\documentclass[aps,prb,preprint,superscriptaddress,showpacs]{revtex4}

\usepackage{textcomp}
\usepackage{amsmath}
\usepackage{amssymb}
\usepackage{graphicx}

\newcommand{\kb}{k_\text{B}}
\newcommand{\avgomega}{\langle \omega \rangle}
\newcommand{\logavgomega}{\omega_\text{ln}}
\newcommand{\jmmapprox}{\sim \! \!}

\begin{document}

\title{High-Temperature Superconductivity in Atomic Metallic Hydrogen}

\author{Jeffrey M. McMahon}
\email[]{mcmahonj@illinois.edu}
\affiliation{Department of Physics, University of Illinois at Urbana-Champaign, Illinois 61801, USA}

\author{David M. Ceperley}
\email[]{ceperley@ncsa.uiuc.edu}
\affiliation{Department of Physics, University of Illinois at Urbana-Champaign, Illinois 61801, USA}
\affiliation{NCSA, University of Illinois at Urbana-Champaign, Illinois 61801, USA}

\date{\today}

\begin{abstract}
Superconductivity in the recently proposed ground-state structures of atomic metallic hydrogen is investigated over the pressure range $500$ GPa to $3.5$ TPa. Near molecular dissociation, the electron--phonon coupling $\lambda$ and renormalized Coulomb repulsion are similar to the molecular phase. A continuous increase in the critical temperature $T_c$ with pressure is therefore expected, to $\jmmapprox 356$K near $500$ GPa. As the atomic phase stabilizes with increasing pressure, $\lambda$ increases, causing $T_c$ to approach $\jmmapprox 481$K near $700$ GPa. At the first atomic--atomic structural phase transformation ($\jmmapprox 1$ -- $1.5$ TPa), a discontinuous jump in $\lambda$ occurs, causing a significant increase in $T_c$ of up to $764$K. 
\end{abstract}

\pacs{74.20.Pq, 74.10.+v, 74.62.Fj, 74.20.Fg}


\maketitle


\section{Introduction}
\label{sec:intro}

At relatively low pressure, hydrogen exists in an insulating molecular phase. In 1935, Wigner and Huntington predicted that sufficient pressure would cause both a molecular-to-atomic transition and metallization \cite{metallic-H_Wigner-JCP-1935}. Recent \textit{ab initio} calculations support these predictions, and have revealed the precise details associated with both effects. Calculations based on \textit{ab initio} random structure searching by Pickard and Needs \cite{AIRSS_H2-phaseIII_Needs-NatPhys-2007} as well as McMahon and Ceperley \cite{McMahon_atomic-H_ground-state_2011} suggest that the molecular-to-atomic transition occurs near $500$ GPa, the latter study also revealing a profusion of structures that atomic hydrogen adopts; and exact-exchange calculations based on density-functional theory (DFT) by St\"{a}dele and Martin \cite{DFT-EXX_H-metallization_Martin-PRL-2000} suggest a metallization pressure of at least $400$ GPa. In 1968, Ashcroft predicted an even further transition in high-pressure hydrogen, a metallic-to-superconducting one \cite{high-Tc_SC_H_Ashcroft-PRL-1968}. Within the framework of Bardeen--Cooper--Schrieffer (BCS) theory \cite{BCS-theory_PR-1957}, three key arguments support this prediction: (i) the ions in the system are single protons, and their small masses cause the vibrational energy scale of the phonons to be remarkably high (e.g., $k_\text{B} \avgomega \approx 2300$K near $500$ GPa -- see below), as is thus the prefactor in the expression for the critical temperature $T_c$; (ii) since the electron--ion interaction is due to the bare Coulomb attraction, the electron--phonon coupling should be strong; and (iii) at the high pressures at and above metallization, the electronic density of states $N(0)$ at the Fermi surface should be large and the Coulomb repulsion between electrons should be low, typical features of high-density systems. These arguments will be revisited, and demonstrated to indeed be the case, below.

Ever since the prediction of high-$T_c$ superconductivity in hydrogen \cite{high-Tc_SC_H_Ashcroft-PRL-1968}, a large number of efforts have focused on determining the precise value(s) of $T_c$ \cite{H_SC_hcp_Stoll-Physica-1971, H_SC_fcc_Caron-PRB-1974, H_SC_Sinha-BookCh-1976, H_SC_Switendick-BookCh-1976, H_SC_Whitmore-PRB-1979, H_SC_fcc_Freeman-PRB-1984, H_SC_dsh_Cohen-Nature-1989, H_SC_9R_Cohen-PRB-1991, H_SC_mu_Ashcroft-PRL-1997, H_SC_fcc-2TPa_Savrasov-SSC-2001, H_SC_fcc-2TPa_Jarosik-SSC-2009, H_SC_I41amd_PoorQual_Liu-PhysLettA-2011, H2_high-Tc_properties_Jarosik-arXiv-2011, H2_Cmca_SC_SC-DFT_Gross-PRL-2008, H2_SC_Cmca_detail1_Gross-PRB-2010, H2_SC_Cmca_detail2_Gross-PRB-2010}. In the molecular phase, the high-pressure metallic $Cmca$ structure (which transitions to the atomic phase \cite{AIRSS_H2-phaseIII_Needs-NatPhys-2007, McMahon_atomic-H_ground-state_2011}) has recently been studied in-depth \cite{H2_Cmca_SC_SC-DFT_Gross-PRL-2008, H2_SC_Cmca_detail1_Gross-PRB-2010, H2_SC_Cmca_detail2_Gross-PRB-2010}, and shown to have a $T_c$ that increases up to $242$K near $450$ GPa. 
In the atomic phase, estimations of $T_c$ have varied widely, but in general suggest a large increase with pressure \cite{H_SC_hcp_Stoll-Physica-1971, H_SC_fcc_Caron-PRB-1974, H_SC_Sinha-BookCh-1976, H_SC_Switendick-BookCh-1976, H_SC_Whitmore-PRB-1979, H_SC_fcc_Freeman-PRB-1984, H_SC_dsh_Cohen-Nature-1989, H_SC_9R_Cohen-PRB-1991, H_SC_mu_Ashcroft-PRL-1997, H_SC_fcc-2TPa_Savrasov-SSC-2001, H_SC_fcc-2TPa_Jarosik-SSC-2009, H_SC_I41amd_PoorQual_Liu-PhysLettA-2011, H2_high-Tc_properties_Jarosik-arXiv-2011}. Early estimates suggested that $T_c \approx 135$ -- $170$K near $400$ GPa (although, it is now believed that this is inside the molecular phase \cite{McMahon_atomic-H_ground-state_2011, AIRSS_H2-phaseIII_Needs-NatPhys-2007}, as discussed above) \cite{H_SC_9R_Cohen-PRB-1991}; near $480$ -- $802$ GPa, more recent estimations suggest that $T_c \approx 282$ -- $291$K \cite{H_SC_I41amd_PoorQual_Liu-PhysLettA-2011}; and near $2$ TPa, calculations suggest that $T_c$ can reach $\jmmapprox 600$ -- $631$K in the face-centered cubic (fcc) lattice \cite{H_SC_fcc-2TPa_Savrasov-SSC-2001, H_SC_fcc-2TPa_Jarosik-SSC-2009}. The latter two studies will be discussed further below.

\begin{figure}
  \includegraphics[scale=0.18, bb=0 0 1308 739]{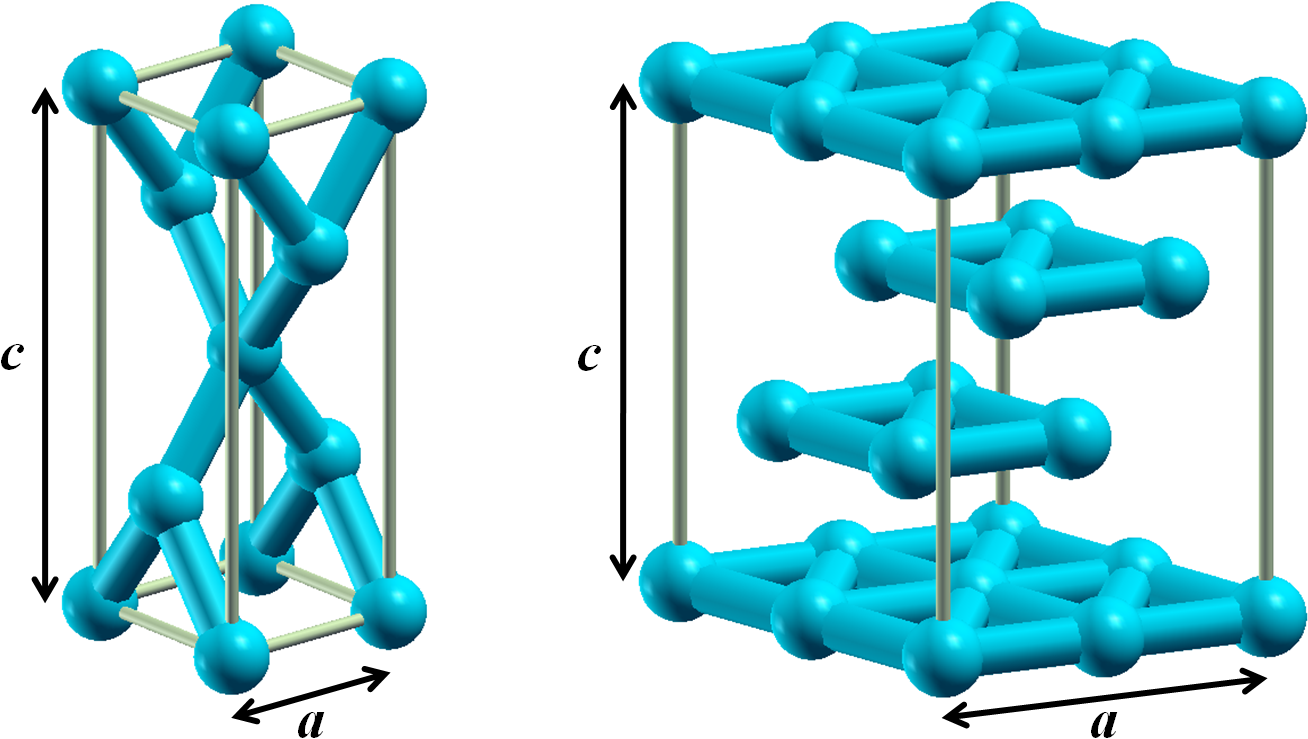}
  \caption{(color online). Ground-state structures of atomic metallic hydrogen. (left) Conventional unit-cell of $I4_1/amd$ at $700$ GPa. (right) $2 \times 2 \times 1$ supercell of $R$-$3m$ at $2$ TPa. $a$ and $c$ parameters are shown in the figure, as discussed in the text. Fictitious bonds have been drawn for clarity.}
  \label{fig:I41amd_R-3m}
\end{figure}

However, previous studies of superconductivity in the atomic phase have simply assumed candidate ground-state structures, in a number of cases the fcc lattice \cite{H_SC_fcc_Caron-PRB-1974, H_SC_Sinha-BookCh-1976, H_SC_Switendick-BookCh-1976, H_SC_fcc_Freeman-PRB-1984, H_SC_fcc-2TPa_Savrasov-SSC-2001, H_SC_fcc-2TPa_Jarosik-SSC-2009}. Recently though, McMahon and Ceperley demonstrated that such structures are incorrect, and provided a comprehensive picture of the (presumably correct) ground-state structures from $500$ GPa to $5$ TPa \cite{McMahon_atomic-H_ground-state_2011}. Molecular hydrogen was shown to dissociate near $500$ GPa, consistent with the predictions of Pickard and Needs \cite{AIRSS_H2-phaseIII_Needs-NatPhys-2007}. With increasing pressure, atomic hydrogen passes through two ground-state structural phases before transforming to a close-packed lattice, such as fcc or possibly the hexagonal close-packed (hcp) lattice. The first is a body-centered tetragonal structure with space-group $I4_1/amd$ (Hermann--Mauguin space-group symbol, international notation) with a $c/a$ ratio greater than unity, as shown in Fig.\ \ref{fig:I41amd_R-3m}.
Including estimates of proton zero-point energies (ZPEs), $I4_1/amd$ was demonstrated to transform into a layered structure with space-group $R$-$3m$ near $1$ TPa, also shown in Fig.\ \ref{fig:I41amd_R-3m}, which is similar to a possible high-pressure phase of lithium \cite{Li_high-P_exp_R-3m_Novikov-Nature-2000}. $R$-$3m$ remains stable to $\jmmapprox 3.5$ TPa, compressing to a close packed lattice. Given such novel crystal phases and that $T_c$ can be very sensitive to structural details \cite{H_SC_Whitmore-PRB-1979}, and that modern methods of calculating values of $T_c$ should be more accurate than those used in earlier studies, it is of great interest to re-investigate the long-outstanding predictions of superconductivity in atomic metallic hydrogen. 

This Article is outlined as follows. In Section \ref{sec:theory}, the theoretical background used for estimating $T_c$ (in this work) is presented. Computational details are given in Section \ref{sec:comp-details}. In Section \ref{sec:properties}, properties of the ground-state structures of atomic metallic hydrogen as a function of pressure, such as lattice parameters and vibrational properties influencing the $I4_1/amd \rightarrow R$-$3m$ transition, are presented and discussed. Superconductivity is investigated in Section \ref{sec:SC}. Section \ref{sec:concl} concludes.

\section{Theoretical Background}
\label{sec:theory}

According to the BCS theory of superconductivity \cite{BCS-theory_PR-1957}, there is a simple relationship between $T_c$, the average phonon frequency $\avgomega$, $N(0)$, and the pairing potential $V$ arising from the electron--phonon interaction,
\begin{equation}
  \label{eq:BCS}
  \kb T_c = 1.14 \avgomega \exp \left[ -\frac{1}{N(0) V} \right]
\end{equation}
where $\kb$ is Boltzmann's constant. This relation is valid as long as $\kb T_c \ll \avgomega$, corresponding to weak coupling -- see below.

McMillan later solved the finite-temperature Eliashberg equations for $T_c$ \cite{McMillan-formula_PR-1968}, which including a correction by Dynes \cite{McMillan-mod_Dynes-SSC-1972} can be written as
\begin{equation}
  \label{eq:McMillan}
  \kb T_c = \frac{\avgomega}{1.2} \exp \left[ -\frac{1.04 \left( 1 + \lambda \right)}{\lambda - \mu^* \left( 1 + 0.62 \lambda \right) } \right] ~~ ,
\end{equation}
where $\lambda$ is the attractive electron--phonon-induced interaction and $\mu^*$ is the renormalized Coulomb repulsion. In high-density atomic hydrogen, Ashcroft \cite{H_SC_mu_Ashcroft-PRL-1997} demonstrated via an \textit{ab initio} calculation that $\mu^* = 0.089$, which is remarkably close to $\mu^* = 0.085$ obtained from the Bennemann--Garland formula \cite{H_SC_fcc-2TPa_Jarosik-SSC-2009}, both results similar to the somewhat standard value for high density systems of $\mu^* \approx 0.1$. In this work, we therefore take $\mu^* = 0.089$ for estimating $T_c$. It should be noted that this approximation fails in molecular hydrogen \cite{H_SC_mu_Ashcroft-PRL-1997}, as investigated thoroughly in Refs.\ \onlinecite{H2_Cmca_SC_SC-DFT_Gross-PRL-2008, H2_SC_Cmca_detail1_Gross-PRB-2010, H2_SC_Cmca_detail2_Gross-PRB-2010} using a specialized formulation of DFT for superconductivity where $\mu^*$ is calculated \textit{ab initio}. Interestingly though, at high densities (e.g., near molecular dissociation) $\mu^*$ is found to nonetheless be $0.08$ for pressures just above $460$ GPa \cite{H2_SC_Cmca_detail2_Gross-PRB-2010}.

For $\lambda \gtrsim 1.3$ (which in fact corresponds to the situations considered herein), Eq.\ (\ref{eq:McMillan}) often provides a lower bound to $T_c$. In this case, both a strong-coupling correction as well as a correction for the shape-dependence of $T_c$ with $\avgomega$ must be made. These corrections will be shown to be especially important in atomic metallic hydrogen, where both $\lambda$ and $\avgomega$ are large. These corrections are both included in the Allen--Dynes equation \cite{Allen-Dynes_McMillan-formula_PRB-1975},
\begin{equation}
  \label{eq:Allen-Dynes}
  \kb T_c = f_1 f_2 \frac{\logavgomega}{1.2} \exp \left[ -\frac{1.04 \left( 1 + \lambda \right)}{\lambda - \mu^* \left( 1 + 0.62 \lambda \right) } \right]
\end{equation}
where $\logavgomega$ is the logarithmic average frequency [i.e., $\ln (\logavgomega) = \langle \ln \omega \rangle$ ] and
\begin{equation}
  \label{eq:f1_AD}
  f_1 = \left[ 1 + \left( \lambda / \Lambda_1 \right)^{3/2} \right]^{1/3}
\end{equation}
\begin{equation}
  \label{eq:f2_AD}
  f_2 = 1 + \frac{ \left( \bar{\omega}_2 / \logavgomega - 1\right) \lambda^2 }{\lambda^2 + \Lambda_2^2}
\end{equation}
denote the strong-coupling and shape corrections, respectively, where $\bar{\omega}_2 = \langle \omega^2 \rangle^{1/2}$ and $\Lambda_1$ and $\Lambda_2$ are fitting parameters (e.g., to full solutions of the Eliashberg equations).

In the original Allen--Dynes equation \cite{Allen-Dynes_McMillan-formula_PRB-1975},
\begin{equation}
  \label{eq:Lambda1_AD}
  \Lambda_1 = 2.46 \left( 1 + 3.8 \mu^* \right)
\end{equation}
\begin{equation}
  \label{eq:Lambda2_AD}
  \Lambda_2 = 1.82 \left( 1 + 6.3 \mu^* \right) \left( \bar{\omega}_2 / \logavgomega \right) ~~ .
\end{equation}
However, a least-squares analysis between $T_c$ as predicted by Eq.\ (\ref{eq:Allen-Dynes}) and that calculated numerically in the Eliashberg formalism for an fcc lattice of atomic metallic hydrogen at $2$ TPa \cite{H_SC_fcc-2TPa_Jarosik-SSC-2009} suggests the following reparametrization
\begin{equation}
  \label{eq:Lambda1_SSC}
  \Lambda_1 = 2.26 \left( 1 - 1.28 \mu^* \right)
\end{equation}
\begin{equation}
  \label{eq:Lambda2_SSC}
  \Lambda_2 = 2.76 \left( 1 + 8.86 \mu^* \right) \left( \bar{\omega}_2 / \logavgomega \right) ~~ ,
\end{equation}
which interestingly provides more accurate values of $T_c$ for a selection of low-temperature superconductors as well \cite{H_SC_fcc-2TPa_Jarosik-SSC-2009}. In passing, we note that there is a very recent further reparametrization by the same authors \cite{H2_high-Tc_properties_Jarosik-arXiv-2011} that appears especially well-suited for calculating $T_c$ for a range of $\mu^*$ values (which could be useful for studying both the molecular and atomic phases concurrently, for example).

In this work, estimates of $T_c$ are made using both Eqs.\ (\ref{eq:McMillan}) and (\ref{eq:Allen-Dynes}) as well as both parametrization for $\Lambda_1$ and $\Lambda_2$ to give a range of values for $T_c$.

\section{Computational Details}
\label{sec:comp-details}

All calculations were performed using the Quantum ESPRESSO \textit{ab initio} DFT code \cite{QE-2009}. A norm-conserving Troullier--Martins pseudopotential \cite{TM-PP_Troullier-Martins-PRB-1991} with a core radius of $0.65$ a.u.\ was used to replace the $1/r$ Coulomb potential of hydrogen, which is sufficiently small to ensure no core-overlap up to the highest pressure considered in this work ($3.5$ TPa). The Perdew-Burke-Ernzerhof exchange and correlation functional \cite{PBE_exch-correl_PRL-1996} was also used for all calculations, as was a basis set of plane-waves with a cutoff of $120$ Ry, giving a convergence in energy to better than $\jmmapprox 0.2$ mRy/proton, and $24^3$ \textbf{k}-points for Brillouin-zone (BZ) sampling. Phonons were calculated using density functional perturbation theory as implemented within Quantum ESPRESSO. Additional computational details pertaining to calculations of the electron--phonon interactions will be provided and discussed in Section \ref{sec:SC}.

\section{Ground-State Structures of Atomic Metallic Hydrogen}
\label{sec:properties}

In this section, we discuss the structural changes that occur in atomic metallic hydrogen as a function of pressure. On the basis of our previous study \cite{McMahon_atomic-H_ground-state_2011}, we consider $I4_1/amd$ at pressures from $500$ GPa to $1.5$ TPa and $R$-$3m$ from $1$ to $3.5$ TPa. We first consider the lattice changes that occur (e.g., compression). We then consider the $I4_1/amd \rightarrow R$-$3m$ transition and discuss the vibrational properties of each structure that contribute to it, in anticipation of the results that are to follow in Section \ref{sec:SC}. A further discussion of the ground-state and metastable structures of atomic metallic hydrogen can be found in Ref.\ \onlinecite{McMahon_atomic-H_ground-state_2011} and a thorough discussion of tetragonal structures of atomic hydrogen (including the $I4_1/amd$ structure) can be found in Ref.\ \onlinecite{H_tet-structs_Matsubara-PRB-1997}.

\subsection{Lattice Parameters}
\label{lb:subsec_struct}

In terms of their primitive unit-cells, $I4_1/amd$ is tetragonal with $a = b \neq c$ with two symmetry inequivalent atoms at Wyckoff positions $(0, 0, 1/2)$ and $(0, 1/2, 3/4)$, and $R$-$3m$ is hexagonal (also with $a = b \neq c$) and a single symmetry inequivalent atom at the origin. The lattice parameters of both structures can therefore be specified completely by $a$ and the $c/a$ ratio, as indicated in Fig.\ \ref{fig:I41amd_R-3m}. For the pressure ranges under consideration, the lattice parameters and corresponding Wigner--Seitz radii $r_s$ are shown in Tables \ref{Tb:a_ca_I41amd} and \ref{Tb:a_ca_R-3m}, respectively.
\begin{table}
  \caption{Lattice parameters and corresponding Wigner--Seitz radii $r_s$ of $I4_1/amd$ as a function of pressure.}
  \label{Tb:a_ca_I41amd}
  \begin{ruledtabular}
    \begin{tabular}{c c c c}
      Pressure (TPa)     & $a$ (a.u.)          & $c/a$                  & $r_s$ (a.u.) \\
      \hline
      \hline
      $0.5$              & $2.299$             & $2.545$                & $1.226$  \\
      \hline
      $0.6$              & $2.227$             & $2.599$                & $1.197$  \\
      \hline
      $0.7$              & $2.134$             & $2.764$                & $1.170$  \\
      \hline
      $0.8$              & $2.094$             & $2.769$                & $1.149$  \\
      \hline
      $0.9$              & $2.058$             & $2.774$                & $1.130$  \\
      \hline
      $1.0$              & $2.027$             & $2.778$                & $1.113$  \\
      \hline
      $1.5$              & $1.893$             & $2.849$                & $1.049$  \\
    \end{tabular}
  \end{ruledtabular}
\end{table}
\begin{table}
  \caption{Lattice parameters and corresponding Wigner--Seitz radii $r_s$ of $R$-$3m$ as a function of pressure.}
  \label{Tb:a_ca_R-3m}
  \begin{ruledtabular}
    \begin{tabular}{c c c c}
      Pressure (TPa)     & $a$ (a.u.)           & $c/a$                  & $r_s$ (a.u.) \\
      \hline
      \hline
      $1.0$              & $1.832$              & $3.236$                & $1.111$  \\
      \hline
      $1.5$              & $1.758$              & $3.061$                & $1.047$  \\
      \hline
      $2.0$              & $1.685$              & $3.054$                & $1.002$  \\
      \hline 
      $2.5$              & $1.629$              & $3.051$                & $0.969$  \\
      \hline
      $3.0$              & $1.584$              & $3.047$                & $0.942$  \\
      \hline
      $3.5$              & $1.564$              & $2.943$                & $0.919$  \\
    \end{tabular}
  \end{ruledtabular}
\end{table}

Between $500$ -- $700$ GPa, $I4_1/amd$ resists compression along the $c$ axis, as can be seen in the $c/a$ ratio which increases from $2.545$ to $2.764$. Above $700$ GPa the resistance continues, but the compression becomes much more uniform. For example, by $1.5$ TPa the $c/a$ ratio increases to only $2.849$. In $R$-$3m$, on the other hand, the $c/a$ ratio remains relatively constant near $3.05$ -- $3.06$. However, near the predicted transition pressures of $\jmmapprox 1$ and $3.5$ TPa (see below and Ref.\ \onlinecite{McMahon_atomic-H_ground-state_2011}) there is a preferred compression along the $c$ axis. In fact, not including the ZPE suggests that $R$-$3m$ continues to compress along the $c$ axis to fcc above $5$ TPa \cite{McMahon_atomic-H_ground-state_2011}.

\subsection{\boldmath $I4_1/amd \rightarrow R$-$3m$ Transition}
\label{lb:subsec_transit}

Static-lattice enthalpy calculations indicate that $I4_1/amd$ transforms to $R$-$3m$ near $2.5$ TPa, but dynamic-lattice calculations (in the harmonic approximation) suggest that this pressure is significantly reduced to $\jmmapprox 1$ TPa \cite{McMahon_atomic-H_ground-state_2011}. 
In this section, we use the harmonic and quasiharmonic approximations to further investigate the $I4_1/amd \rightarrow R$-$3m$ transition, in anticipation of the results that are to follow in Section \ref{sec:SC}.

Ground-state enthalpies for $I4_1/amd$ and $R$-$3m$ (defined by the parameters in Tables \ref{Tb:a_ca_I41amd} and \ref{Tb:a_ca_R-3m}) were calculated at $1$ and $1.5$ TPa; Table \ref{Tb:gs_H_ZPE}.
\begin{table}
  \caption{Ground-state enthalpies and zero-point energies of $I4_1/amd$ and $R$-$3m$ at $1$ and $1.5$ TPa. Pressures $P$ are in TPa and enthalpies and energies are in Ry/proton.}
  \label{Tb:gs_H_ZPE}
  \begin{ruledtabular}
    \begin{tabular}{c c c}
                              & $P = 1.0$     & $1.5$       \\
      \hline
      \hline
      $H$                     &      &        \\
      \hline
      \hline
      $I4_1/amd$              & $-0.49955$     & $-0.32022$   \\
      \hline
      $R$-$3m$                & $-0.49534$     & $-0.31768$   \\
      \hline
      \hline
      $E_\text{ZPE}$          &      &        \\
      \hline
      \hline
      $I4_1/amd$              & $0.02708$     & $0.03120$   \\
      \hline
      $R$-$3m$                & $0.02395$     & $0.02769$   \\
      \hline
      \hline
      $H + E_\text{ZPE}$      &      &        \\
      \hline
      \hline
      $I4_1/amd$              & $-0.47247$     & $-0.28902$   \\
      \hline
      $R$-$3m$                & $-0.47140$     & $-0.28999$   \\
    \end{tabular}
  \end{ruledtabular}
\end{table}
ZPEs at each pressure were estimated using the harmonic approximation: $E_\text{ZPE} = \int d\omega ~ F(\omega) \hbar \omega / 2$, where $F(\omega)$ is the phonon density of states (PHDOS), and are shown in Table \ref{Tb:gs_H_ZPE} as well. Neglecting zero-point pressures and making the simple approximation that the total enthalpies are given by $H + E_\text{ZPE}$ (as was done in Ref.\ \onlinecite{McMahon_atomic-H_ground-state_2011}) suggests that the $I4_1/amd \rightarrow R$-$3m$ transition occurs nearly midway between $1$ and $1.5$ TPa (see Table \ref{Tb:gs_H_ZPE}), which is very close to, but slightly higher than our original estimate of $1$ TPa \cite{McMahon_atomic-H_ground-state_2011}. Going beyond this approximation, the total enthalpies, including the zero-point pressures, can be estimated using a linear approximation,
\begin{equation}
  \label{eq:Htot_ZPE}
  H_\text{tot} = H_\text{avg} + E_\text{ZPE, avg} + p_\text{ZPE} V_\text{avg}
\end{equation}
where
\begin{equation}
  \label{eq:p_ZPE}
  p_\text{ZPE} = - \frac{\partial E_\text{ZPE}}{\partial V}
\end{equation}
is the zero-point pressure, $V$ is the volume, and the subscripts avg denote the average values of each quantity between $1$ and $1.5$ TPa. Note that the latter two quantities in Eq.\ (\ref{eq:Htot_ZPE}) correspond to the zero-point enthalpy. Estimating $p_\text{ZPE}$ using a simple finite-difference gives total enthalpies of $-0.35765$ and $-0.35976$ Ry/proton for $I4_1/amd$ and $R$-$3m$, respectively. This suggests that the actual transition pressure is a bit lower than the simple enthalpy estimate, and is in fact in agreement with our original prediction of $\jmmapprox 1$ TPa \cite{McMahon_atomic-H_ground-state_2011}.

\begin{figure}
  \includegraphics[scale=0.27, bb=0 0 905 658]{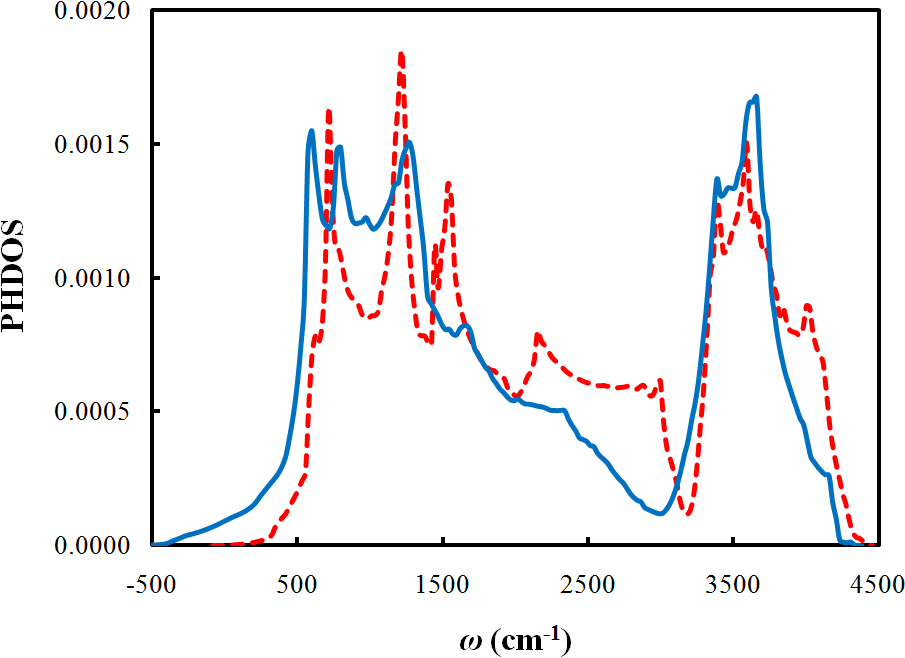}
  \caption{(color online). PHDOS of $I4_1/amd$ (dashed red line) and $R$-$3m$ (solid blue line) at $1.5$ TPa. The results have been normalized by the number of atoms per primitive unit-cell. Negative values indicate imaginary frequencies.}
  \label{fig:phdos_I41amd_R-3m}
\end{figure}

As can be inferred from Table \ref{Tb:gs_H_ZPE} and the discussion above, the large decrease in the $I4_1/amd \rightarrow R$-$3m$ transition pressure from the static-lattice prediction ($\jmmapprox 2.5$ TPa \cite{McMahon_atomic-H_ground-state_2011}) arises primarily from the significantly lower $E_\text{ZPE}$ in $R$-$3m$, as well as a more minor contribution from the lower $p_\text{ZPE}$.
To help understand this, the PHDOS for both structures is shown in Fig.\ \ref{fig:phdos_I41amd_R-3m}.
%
It can be seen that there are three differences that lead to this behavior: (i) the density of high-frequency phonons is greater for $I4_1/amd$, and also occurs at higher frequencies ($3180$ -- $4430$ cm$^{-1}$ vs $3000$ -- $4230$ cm$^{-1}$); (ii) $I4_1/amd$ has a significant density of mid-frequency phonons ($\jmmapprox 1400$ -- $3000$ cm$^{-1}$), while such modes are mostly absent in $R$-$3m$ (e.g., $I4_1/amd$ shows significant peaks at $1510$, $2150$, and $2990$ cm$^{-1}$); and therefore (iii) the PHDOS for $R$-$3m$ is mostly concentrated at low frequencies ($\lesssim 1400$ cm$^{-1}$). 

In passing, we note that $R$-$3m$ shows a small density of imaginary phonon states at $1.5$ TPa. However, estimating the resulting energy within the harmonic approximation \cite{McMahon_atomic-H_ground-state_2011} shows that it only integrates to $1.372 \cdot 10^{-5}$ Ry/proton. While this is within the accuracy of our calculations, this behavior is in fact expected considering that it is indicative of instability in a lattice of ions treated classically; and classically, the $I4_1/amd \rightarrow R$-$3m$ transition occurs near $2.5$ TPa \cite{McMahon_atomic-H_ground-state_2011}, as discussed above. This is further confirmed by the fact that the instability goes to zero with increasing pressure, while such behavior begins to develop in $I4_1/amd$ -- see Ref.\ \onlinecite{McMahon_atomic-H_ground-state_2011}.

Considering that the PHDOSs are quite different between $I4_1/amd$ and $R$-$3m$ and it is finite-temperature effects that are focused on below (i.e., $T_c$), the possibility of vibrational entropic stabilization of one phase over the other exists. In order to estimate this, the quasiharmonic approximation can be used,
\begin{equation}
  \label{eq:quasiharm}
  F(V,T) = E_0(V) + k_\text{B} T \int_0^\infty d\omega ~ F(\omega) \ln \left[ \sinh \left( \frac{\hbar \omega}{2 k_\text{B} T} \right) \right]
\end{equation}
where $F(V,T)$ is the Helmholtz free energy at volume $V$ and temperature $T$ and $E_0(V)$ is the static-lattice energy. From this, the Gibbs free energy $G$ can be calculated via $G = F + pV$, given the pressure $p$. At $T = 0$K, $p$ is given by the external pressure plus the zero-point pressure [Eq.\ (\ref{eq:p_ZPE})]. However, for a fixed $V$, $p$ is actually a function of $T$, due to thermal expansion of the lattice caused by anharmonic phonons. Contrary to the expectation that such effects may be large \cite{atomic-H_cryst-struct_Ceperley-PRL-1993}, calculations of the melting line of hydrogen (not shown) \cite{McMahon_H_melting_classical_2011} indicate that in fact thermal expansion is in fact small, at least up to a few hundred K where atomic metallic hydrogen is likely to melt anyway; and since the purpose of this discussion is just to understand qualitative changes that may arise at finite-$T$, we can estimate $p$ using the $T = 0$K value.

\begin{figure}
  \includegraphics[scale=0.27, bb=0 0 908 658]{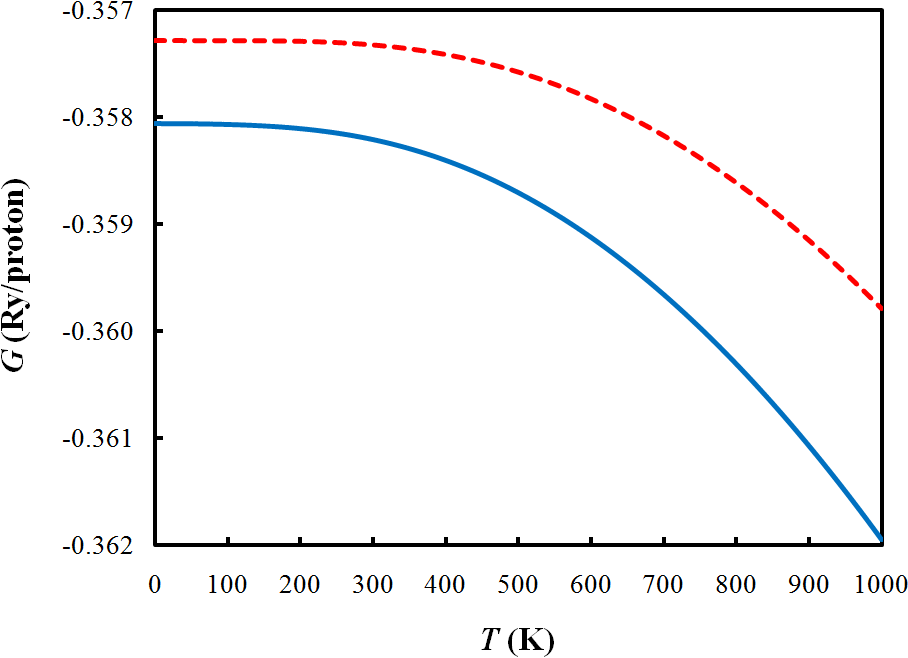}
  \caption{(color online). Gibbs free energy vs temperature for $I4_1/amd$ (dashed red line) and $R$-$3m$ (solid blue line) at $1.5$ TPa.}
  \label{fig:G_I41amd_R-3m}
\end{figure}

Figure \ref{fig:G_I41amd_R-3m} shows the resulting free energy $G$ estimated using Eq.\ (\ref{eq:quasiharm}) and the value of $p$ at $T = 0$K.
Despite the remarkably different PHDOSs (see again Fig.\ \ref{fig:phdos_I41amd_R-3m}), the behaviors of $G$ with $T$ are rather similar for both structures. Thus, temperature is not expected to significantly affect the $I4_1/amd \rightarrow R$-$3m$ transition. 

Based on these results, below we consider $I4_1/amd$ from $500$ GPa to $1.5$ TPa and $R$-$3m$ from $1$ to $3.5$ TPa, and the $I4_1/amd \rightarrow R$-$3m$ transition to occur between $1$ and $1.5$ TPa.

\section{Superconductivity}
\label{sec:SC}

In this section, we investigate superconductivity in the $I4_1/amd$ and $R$-$3m$ structures of atomic metallic hydrogen. We first provide relevant computational details not discussed in Section \ref{sec:comp-details}, and convergence of the parameters necessary to evaluate Eqs.\ (\ref{eq:McMillan}) and (\ref{eq:Allen-Dynes}) with respect to \textbf{q}-points. We then present and discuss the calculated parameters, and finally use them to calculate $T_c$ values.

\subsection{Computational Details}

In order to estimate $T_c$ using Eqs.\ (\ref{eq:McMillan}) and (\ref{eq:Allen-Dynes}), $\avgomega$, $\logavgomega$, $\bar{\omega}_2$, and $\lambda$ must all be determined. Of course, the frequency parameters can be calculated directly from the PHDOS. For example, $\avgomega = (1 / n_\text{ph}) \int d\omega ~ F(\omega) \omega$, where $n_\text{ph}$ is the number of phonon modes and $\int d\omega ~ F(\omega) = n_\text{ph}$. In order to calculate $\lambda$, a (slowly convergent) double-delta integration must be performed on the Fermi surface -- see Ref.\ \onlinecite{SC_implementation_PWSCF_Giannozzi-arXiv-2011} for a complete discussion and the precise implementation details within Quantum ESPRESSO. In order to accurately perform this integration, very dense \textbf{k}-point and \textbf{q}-point grids must be used. For the $I4_1/amd$ and $R$-$3m$ structures, we found that an electronic grid of $48^3$ \textbf{k}-points (and using $24^3$ \textbf{k}-points to calculate phonons, as discussed in Section \ref{sec:comp-details}) gave convergence with no discernible error. 

In order to determine a sufficient density for the \textbf{q}-point grid, we performed a series of calculations with $1^3$, $2^3$, $4^3$, $6^3$, and $8^3$ \textbf{q}-points, using $I4_1/amd$ at $500$ GPa as a test case (we also considered $R$-$3m$ at $2$ TPa -- not shown). It should be kept in mind that such rigorous testing with respect to \textbf{q}-points is especially important in atomic metallic hydrogen, as inadequate sampling has been shown to cause significantly incorrect results \cite{H_SC_dsh_Cohen-Nature-1989, H_SC_9R_Cohen-PRB-1991}. In fact, our calculations below suggest that the results of a recent study considering Cs-IV (which also has the $I4_1/amd$ structure) over a more narrow pressure range than considered here using only $3^3$ \textbf{q}-points \cite{H_SC_I41amd_PoorQual_Liu-PhysLettA-2011} gives somewhat incorrect values for $\lambda$, both in magnitude and trend with pressure. The values of $\lambda$ for the various densities of \textbf{q}-points, as well as values of $\avgomega$ and $\logavgomega$, are shown in Table \ref{Tb:qpt_convg}.
%
%
\begin{table}
  \caption{Convergence of $\avgomega$, $\logavgomega$, and $\lambda$ with the number of \textbf{q}-points for $I4_1/amd$ at $500$ GPa.}
  \label{Tb:qpt_convg}
  \begin{ruledtabular}
    \begin{tabular}{c c c c}
      No.\ of \textbf{q}-points     & $\avgomega$ (K)    & $\logavgomega$ (K)    & $\lambda$ \\
      \hline
      \hline
      $1^3$                         & $1660$             & $1438$                & $17.91$ \\
      \hline
      $2^3$                         & $2307$             & $1953$                & $2.82$ \\
      \hline
      $4^3$                         & $2277$             & $2031$                & $2.06$ \\
      \hline
      $6^3$                         & $2287$             & $1997$                & $1.67$ \\
      \hline
      $8^3$                         & $2295$             & $2068$                & $1.81$ \\
    \end{tabular}
  \end{ruledtabular}
\end{table}
%
Relative convergence in $\lambda$ is seen to require at least $6^3$ \textbf{q}-points (to be within $10\%$ of the converged value, for example). This is likely due to Fermi surface ``hot spots'' that have been shown to exist in other alkali metals \cite{Li_SC_Pickett-PRL-2006}, which can significantly contribute to the electron--phonon interaction. Table \ref{Tb:qpt_convg} also shows, on the other hand, that $\avgomega$ and $\logavgomega$ achieve relative convergence with as little as $2^3$ \textbf{q}-points, which is consistent with the density found necessary in our previous work to accurately calculate the ZPEs of the structures of atomic metallic hydrogen \cite{McMahon_atomic-H_ground-state_2011}. Herein, $8^3$ \textbf{q}-points were used for all calculations (including those in Section \ref{sec:properties}), corresponding to $59$ and $150$ total \textbf{q}-points in the irreducible BZ for $I4_1/amd$ and $R$-$3m$, respectively.

\subsection{Superconducting Parameters}

\begin{figure}
  \includegraphics[scale=0.27, bb=0 0 880 655]{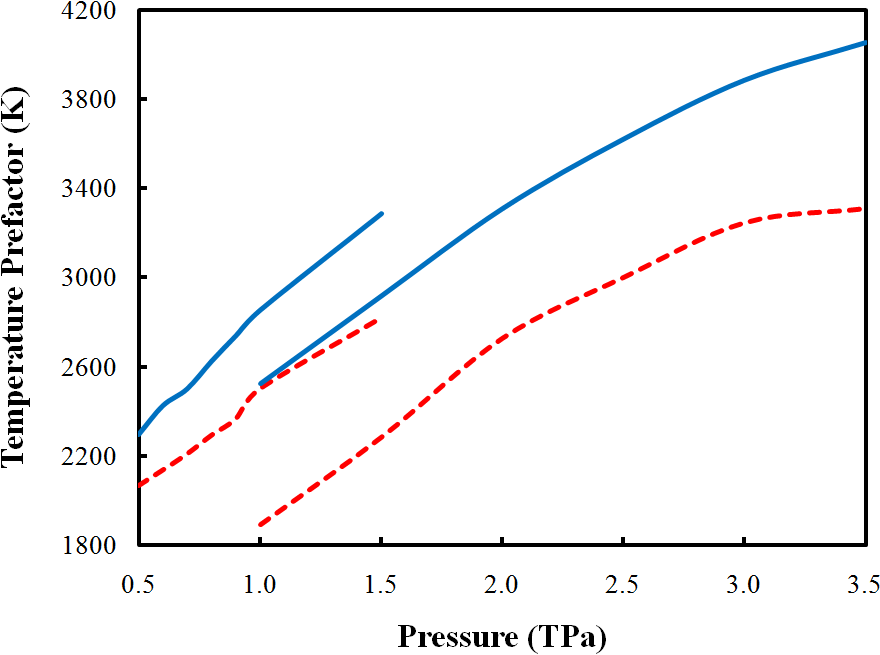}
  \caption{(color online). Temperature prefactors $\avgomega$ (solid blue line) and $\logavgomega$ (dashed red line) as a function of pressure in atomic metallic hydrogen.}
  \label{fig:omega}
\end{figure}

As shown in Fig.\ \ref{fig:omega}, $\avgomega$ and $\logavgomega$ are both extremely high, and increase significantly with pressure \cite{McMahon_atomic-H_ground-state_2011}; $\avgomega$ increases from $2295$K to $4056$K as the pressure is increased from $500$ GPa to $3.5$ TPa, while $\logavgomega$ is significantly less (especially for $R$-$3m$), increasing from $2068$K to $3308$K over the same range. 
%
Furthermore, there is a significant decrease in both $\avgomega$ and $\logavgomega$ at the $I4_1/amd \rightarrow R$-$3m$ transition (e.g., by $765$K and $926$K, respectively, at $1.5$ TPa), consistent with the results and discussion in Section \ref{lb:subsec_transit}.

\begin{figure}
  \includegraphics[scale=0.28, bb=0 0 854 657]{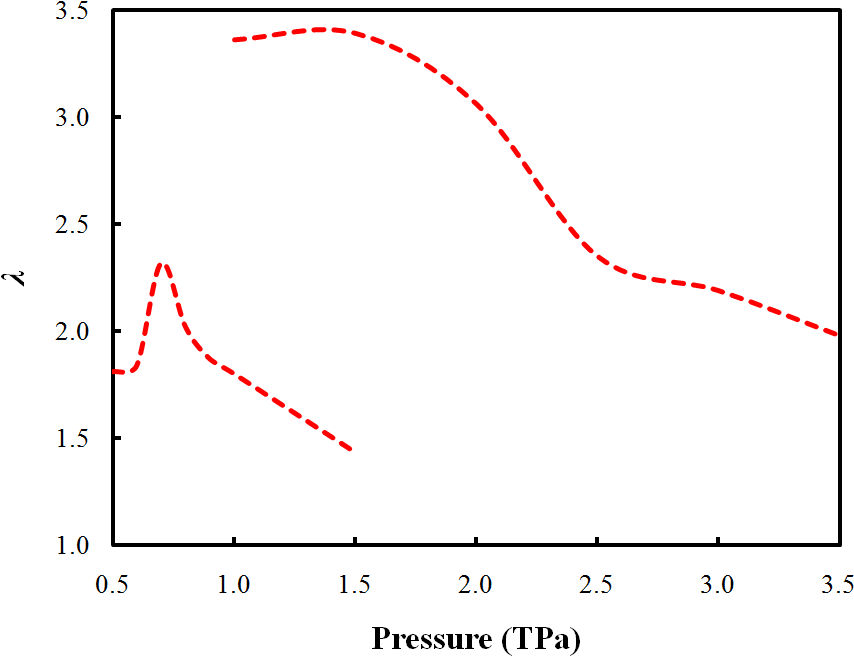}
  \caption{(color online). Electron--phonon-induced interaction $\lambda$ as a function of pressure in atomic metallic hydrogen.}
  \label{fig:lambda}
\end{figure}

More interesting is the behavior of $\lambda$ with pressure; Fig.\ \ref{fig:lambda}. 
%
Near molecular dissociation ($\jmmapprox 500$ GPa), the values of $\lambda$ in both the atomic and molecular phases are remarkably close. In $I4_1/amd$, $\lambda \approx 1.81$ (see also Table \ref{Tb:qpt_convg}), whereas in the molecular phase ($Cmca$) $\lambda \approx 2$ just above $460$ GPa, but appears to slowly decrease with increasing pressure -- see Refs.\ \onlinecite{H2_Cmca_SC_SC-DFT_Gross-PRL-2008, H2_SC_Cmca_detail2_Gross-PRB-2010}. 
Thus, given that $\lambda$ and $\mu^*$ are similar in both phases near molecular dissociation (see again Section \ref{sec:theory} for a discussion of $\mu^*$), a smooth variation in $T_c$ is likely to occur with increasing pressure in this range.

\begin{figure}
  \includegraphics[scale=0.28, bb=0 0 872 658]{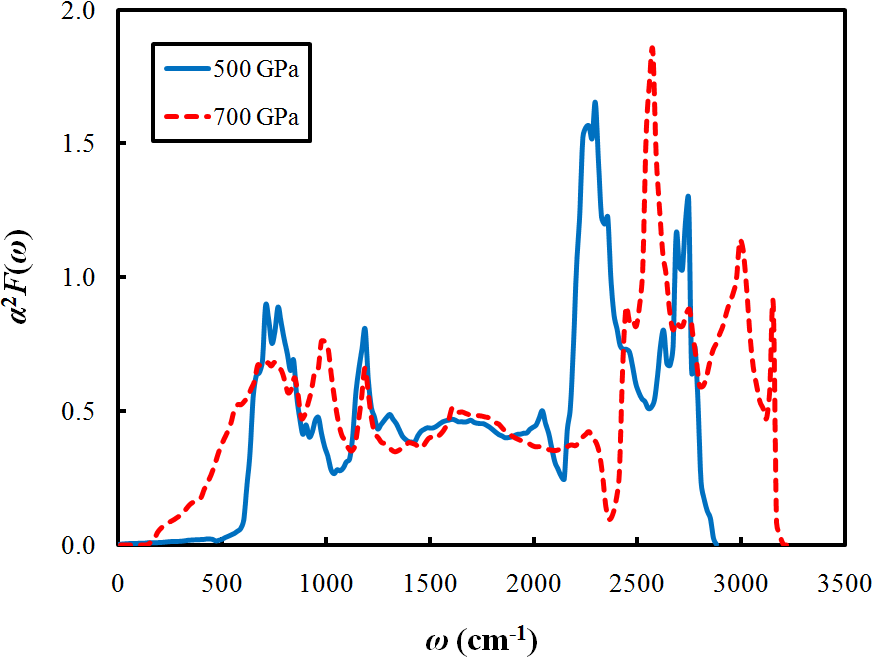}
  \caption{(color online). Electron--phonon spectral function $\alpha^2 F(\omega)$ of $I4_1/amd$ at $500$ and $700$ GPa.}
  \label{fig:a2F_I41amd_500-700GPa}
\end{figure}

A large increase in $\lambda$ is seen to occur from $500$ -- $700$ GPa, from $1.81$ to $2.32$. 
To help understand this, the electron--phonon spectral function, $\alpha^2 F(\omega)$, at $500$ GPa is compared to that at $700$ GPa in Fig.\ \ref{fig:a2F_I41amd_500-700GPa}.
%
It can be seen that there is an increase in coupling to both the low- and high-frequency phonon modes as the atomic phase stabilizes, while there is relatively little change in the coupling to those at mid frequency. The former increase is unexpected, as with increasing pressure the PHDOS shifts to higher frequencies, as is indicated in Fig.\ \ref{fig:omega}. The sharp increase in $\lambda$, along with the increased $\avgomega$ and $\logavgomega$ (see again Fig.\ \ref{fig:omega}), suggests that a correspondingly large increase in $T_c$ should occur over this small pressure range, which is shown below to indeed be the case.

\begin{figure}
  \includegraphics[scale=0.28, bb=0 0 871 659]{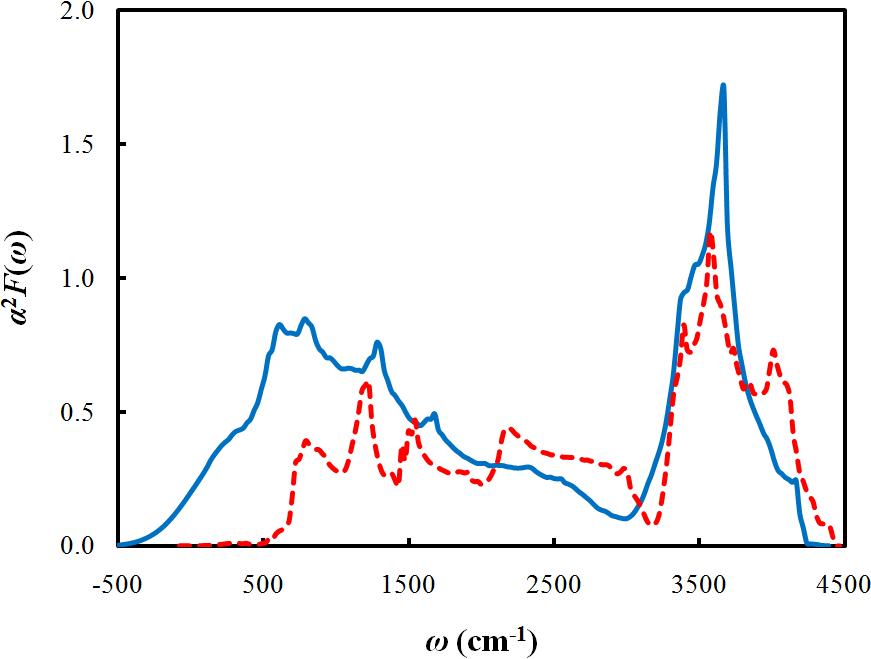}
  \caption{(color online). Electron--phonon spectral function $\alpha^2 F(\omega)$ of $I4_1/amd$ (dashed red line) and $R$-$3m$ (solid blue line) at $1.5$ TPa.}
  \label{fig:a2F_I41amd_R-3m_1p5TPa}
\end{figure}

At the $I4_1/amd \rightarrow R$-$3m$ transition near $1.5$ TPa, a large jump in $\lambda$ occurs, from $1.43$ to $3.39$. This can be understood by comparing $\alpha^2 F(\omega)$ for both structures; Fig.\ \ref{fig:a2F_I41amd_R-3m_1p5TPa}.
In $R$-$3m$, the large value of $\lambda$ is seen to occur from a strong coupling into the low-frequency modes [$\lambda = 2 \int d\omega ~ \alpha^2 F(\omega) / \omega$]. This appears to be due to the correspondingly high PHDOS at low frequencies, which is absent in $I4_1/amd$ (see again Section \ref{lb:subsec_transit}). Comparing Figs.\ \ref{fig:a2F_I41amd_500-700GPa} and \ref{fig:a2F_I41amd_R-3m_1p5TPa} also shows that in $I4_1/amd$ there is decreased coupling into all modes with an increase in pressure above $700$ GPa, especially at low frequencies. 

\begin{figure}
  \includegraphics[scale=0.28, bb=0 0 848 658]{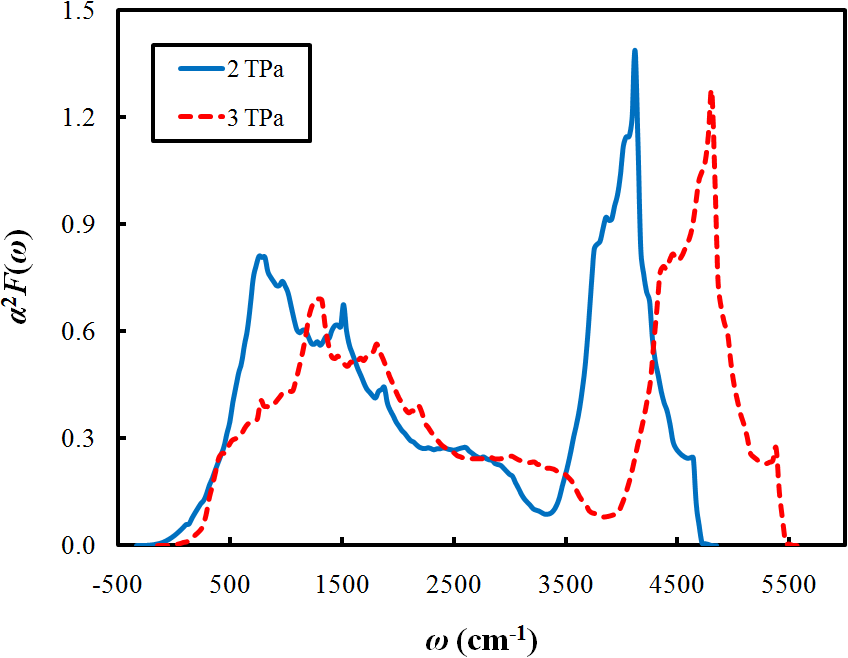}
  \caption{(color online). Electron--phonon spectral function $\alpha^2 F(\omega)$ of $R$-$3m$ at $2$ and $3$ TPa.}
  \label{fig:a2F_R-3m_2-3TPa}
\end{figure}

With increasing pressure, $\lambda$ in $R$-$3m$ decreases from its maximum to $\jmmapprox 1.98$ by $3.5$ TPa. Figure \ref{fig:a2F_R-3m_2-3TPa} shows that this results from a weakened coupling into the low-frequency modes that was responsible for the sharp increase in $\lambda$ in the first place (near the $I4_1/amd \rightarrow R$-$3m$ transition).
This is likely due to an overall decrease in the PHDOS at low frequencies with increasing pressure (not shown). These results, combined with those above, indicates that coupling into the low-frequency modes are the key to achieving a large value of $\lambda$ in atomic metallic hydrogen.

The $\lambda$ values presented above are much less than for the (unstable) fcc lattice. For example, at $2$ TPa $\lambda \approx 3.06$ compared to $\lambda \approx 7$ -- $7.32$ \cite{H_SC_fcc-2TPa_Savrasov-SSC-2001, H_SC_fcc-2TPa_Jarosik-SSC-2009}. The large difference can be attributed to the even higher PHDOS at low frequencies in fcc compared to $R$-$3m$ \cite{McMahon_atomic-H_ground-state_2011}, which was above suggested to lead to very strong electron--phonon coupling. In Ref.\ \onlinecite{H_SC_fcc-2TPa_Savrasov-SSC-2001}, the strong coupling into the low-frequency modes in fcc was attributed to the lattice being close to instability. While this is consistent with (and likely influential) in the strong coupling into $R$-$3m$, which is also close to lattice instability near $1$ -- $1.5$ TPa with protons treated classically (see Section \ref{lb:subsec_transit} and Ref.\ \cite{McMahon_atomic-H_ground-state_2011}), this is not necessarily the cause. For example, such behavior does not always occur, as $\lambda$ for $I4_1/amd$ appears low near its pressure limits of lattice instability (e.g., $\jmmapprox 500$ GPa), while it becomes largest near the center of this range (e.g., $\jmmapprox 700$ GPa) -- see above. 

\subsection{\boldmath $T_c$}

\begin{figure}
  \includegraphics[scale=0.23, bb=0 0 1041 785]{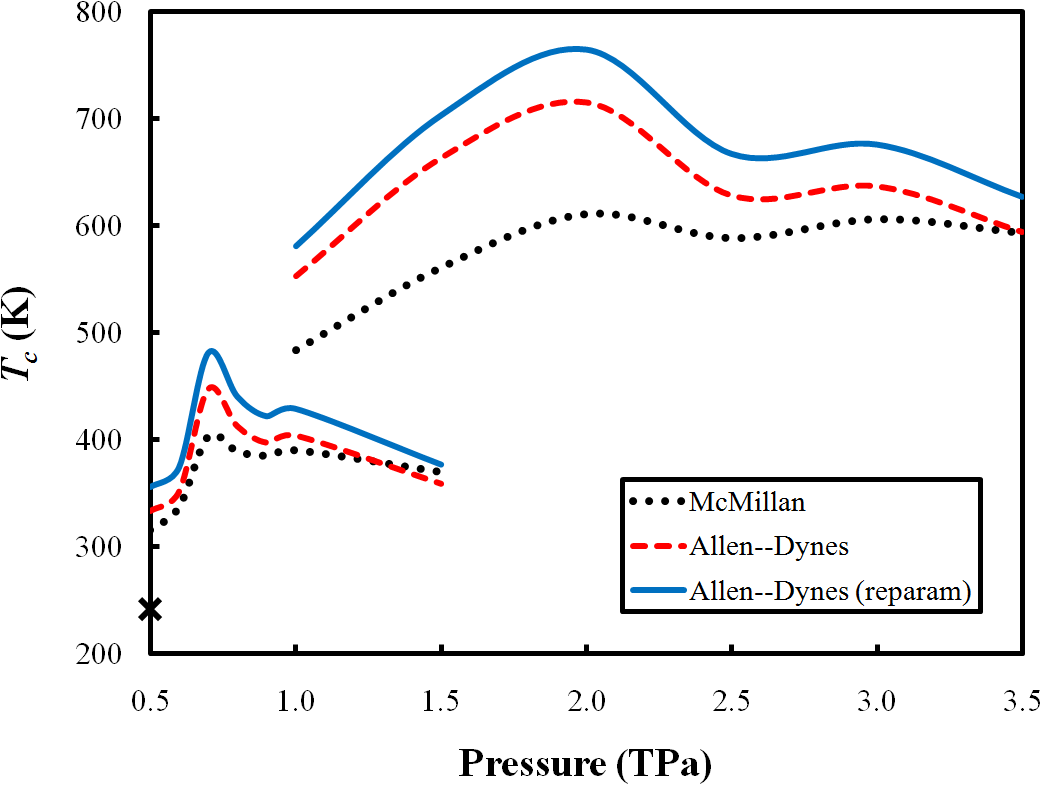}
  \caption{(color online). Values of $T_c$ for atomic metallic hydrogen calculated using Eqs.\ (\ref{eq:McMillan}) and (\ref{eq:Allen-Dynes}). $T_c$ for the high-pressure molecular phase is shown using a symbol.}
  \label{fig:Tc}
\end{figure}

Using the parameters in Figs.\ \ref{fig:omega} and \ref{fig:lambda}, Eqs.\ (\ref{eq:McMillan}) and (\ref{eq:Allen-Dynes}) were used to calculate $T_c$; Fig.\ \ref{fig:Tc}.
%
The values are seen to be remarkably high, but nonetheless consistent with the discussion above. The Allen--Dynes equation and its reparametrization \cite{H_SC_fcc-2TPa_Jarosik-SSC-2009}, in most cases, give much higher estimates than the McMillan formula (as expected based on the discussion in Section \ref{sec:theory}). Given that $\logavgomega$ is significantly less than $\avgomega$, the increase is thus due entirely to the correction factors $f_1$ and $f_2$. Comparing these, in a number of cases, shows that it is $f_1$ (the strong-coupling correction) that is most important, especially in the reparametrized Allen--Dynes equation \cite{H_SC_fcc-2TPa_Jarosik-SSC-2009}. For example, at $700$ GPa $f_1 \approx 1.31$ and $f_2 \approx 1.03$.

Just above molecular dissociation, $T_c \approx 315$ -- $356$K. The increase in $\lambda$ combined with increases in $\avgomega$ and $\logavgomega$ with pressure cause $T_c$ to increase up to $403$ -- $481$K by $700$K. With increasing pressure, $T_c$ then decreases (in the $I4_1/amd$ phase). However, at the $I4_1/amd \rightarrow R$-$3m$ transition, a large jump in $T_c$ then occurs, from $370$ -- $377$K to $561$ -- $703$K. This is due entirely to the jump in $\lambda$, considering that $\avgomega$ and $\logavgomega$ are significantly less in $R$-$3m$ (see Fig.\ \ref{fig:omega}). 
Although, with increasing pressure, $T_c$ then decreases. Thus, $\jmmapprox 764$K represents an approximate upper bound to $T_c$ in atomic metallic hydrogen, and possibly conventional superconductors (i.e., those described by BCS theory) in general. It is interesting to note that secondary maxima in $T_c$ occur in both $I4_1/amd$ and $R$-$3m$. Given that there appears to be monotonic decreases in $\lambda$ above their maxima in both structures (see Fig.\ \ref{fig:lambda}), this behavior is simply due to an interplay between this and $\avgomega$ or $\logavgomega$.

\section{Conclusions}
\label{sec:concl}

In conclusion, we investigated superconductivity in the ground-state structures of atomic metallic hydrogen over the range $500$ GPa to $3.5$ TPa. Near molecular dissociation, the electron--phonon coupling $\lambda$ and renormalized Coulomb repulsion in the atomic phase were demonstrated to be similar to the values of the molecular phase. This suggests a continuous increase in $T_c$ with pressure during the molecular-to-atomic transition, to $\jmmapprox 356$K near $500$ GPa. As the atomic phase stabilizes with increasing pressure, $\lambda$ increases causing $T_c$ to increase to $\jmmapprox 481$K near $700$ GPa. Near the first atomic--atomic structural phase transformation near $1.5$ TPa, a large jump in $\lambda$ occurs due to a high PHDOS at low frequencies, increasing $T_c$ to as high as $764$K. 

While the $T_c$ values presented incredibly high, they are nonetheless reasonable. However, there are two caveats. First of all, even the lowest pressures considered in this work are higher than those currently obtainable experimentally ($342$ GPa \cite{solid_H_exp_highest-P_Ruoff-Nature-1998}). 
Nonetheless, all of them are important to planetary physics (albeit likely at temperatures even higher than the values of $T_c$). The other caveat is that it is quite possible that the $T_c$ values are higher than the melting temperatures of the phases of atomic metallic hydrogen. However, this suggests the interesting possibility that the atomic metallic solid phase of hydrogen (at least the $I4_1/amd$ and $R$-$3m$ structures) may exist entirely in superconducting states.


\begin{acknowledgments}
J.\ M.\ M.\ and D.\ M.\ C.\ were supported by DOE DE-FC02-06ER25794 and DE-FG52-09NA29456. This research was also supported in part by the National Science Foundation through TeraGrid resources provided by NICS under grant number TG-MCA93S030.
\end{acknowledgments}


\end{document}